\documentclass[pra,twocolumn,superscriptaddress]{revtex4-2}

\usepackage{graphicx}
\usepackage{amsfonts}
\usepackage{mathtools}
\usepackage{dsfont}
\usepackage{float}
\usepackage{graphicx}
\usepackage{amssymb}
\usepackage{color}
\usepackage{enumerate}
\usepackage{amsmath}
\usepackage{amstext}
\usepackage{latexsym}
\usepackage{textcomp}
\usepackage[usenames,dvipsnames]{xcolor}
\usepackage[scaled=1]{helvet}
\usepackage{lipsum}
\usepackage{hyperref}
\hypersetup{colorlinks=true,citecolor=Blue,linkcolor=RubineRed,urlcolor=Blue}

\def\la{\langle}
\def\ra{\rangle}

\def\tr{{\rm tr}}

\newcommand{\beq}{\begin{equation}}
\newcommand{\eeq}{\end{equation}}
\newcommand{\beqa}{\begin{eqnarray}}
\newcommand{\eeqa}{\end{eqnarray}}

\begin{document}

\title{Unitarity breaking in self-averaging spectral form factors}

\author{Apollonas S. Matsoukas-Roubeas \href{https://orcid.org/0000-0001-5517-0224}{\includegraphics[scale=0.04]{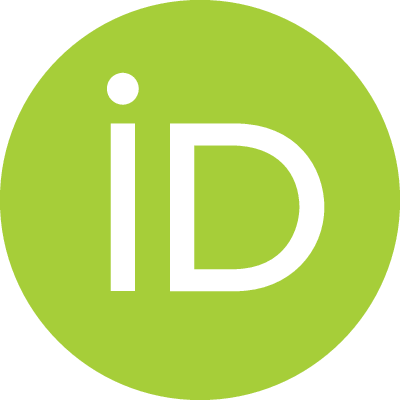}}}
\affiliation{Department of Physics and Materials Science, University of Luxembourg, L-1511 Luxembourg, G. D. Luxembourg}
\author{Mathieu Beau \href{https://orcid.org/0000-0003-0502-6210}{\includegraphics[scale=0.05]{orcidid.eps}}}
\affiliation{Department of Physics, University of Massachusetts, Boston, MA 02125, USA}
\author{Lea F. Santos  \href{https://orcid.org/0000-0001-9400-2709}{\includegraphics[scale=0.04]{orcidid.eps}}}
\affiliation{Department of Physics, University of Connecticut, Storrs, Connecticut 06269, USA}
\author{Adolfo del Campo \href{https://orcid.org/0000-0003-2219-2851}{\includegraphics[scale=0.04]{orcidid.eps}}}
\affiliation{Department of Physics and Materials Science, University of Luxembourg, L-1511 Luxembourg, G. D. Luxembourg}
\affiliation{Donostia International Physics Center, E-20018 San Sebasti\'an, Spain}

\begin{abstract}
The complex Fourier transform of the two-point correlator of the energy spectrum of a quantum system is known as the spectral form factor (SFF). It constitutes an essential diagnostic tool for phases of matter and quantum chaos. In black hole physics, it describes the survival probability (fidelity) of a thermofield double state under unitary time evolution. However, detailed properties of the SFF of isolated quantum systems with generic spectra are smeared out by large temporal fluctuations, whose minimization requires disorder or time averages. This requirement holds for any system size, that is, the SFF is non-self averaging. 
Exploiting the fidelity-based interpretation of this quantity, we prove that using filters, disorder and time averages of the SFF involve unitarity breaking, i.e., open quantum dynamics described by a  quantum channel that suppresses quantum noise. Specifically,  averaging over Hamiltonian ensembles, time averaging, and frequency filters can be described by the class of mixed-unitary quantum channels in which information loss can be recovered.  Frequency filters are associated with a time-continuous master equation generalizing energy dephasing. We also discuss the use of  eigenvalue filters. They are linked to non-Hermitian Hamiltonian evolution without quantum jumps, whose long-time behavior is described by a Hamiltonian deformation. We show that frequency and energy filters make the SFF self-averaging at long times.
\end{abstract}

\maketitle

The spectral form factor (SFF) is an essential diagnostic tool in the characterization of complex quantum systems \cite{Leviandier86,WilkieBrumer91,Alhassid1992,Alhassid93,Haake,MethaBook}.
Given a Hamiltonian $H$ of a single system with spectrum ${\rm Sp}(H)=\{E_n | n=1,\dots,d\}$, the SFF is a real-valued function defined as
\beqa
\label{SFFdef}
{\rm SFF}(t)&=&\left|\frac{Z(\beta+it)}{Z(\beta)}\right|^2\\
&=&\frac{1}{Z(\beta)^2} \sum_{n,m=1}^d e^{-\beta (E_n+E_m)-it(E_n-E_m)},\nonumber
\eeqa
where we use units with $\hbar=1$. The partition function $Z(\beta)=\tr[\exp(-\beta H)]$  is included as a normalization such that ${\rm SFF}(0)=1$. 
Finite values of the inverse temperature $\beta$ exponentially suppress the contribution from   the excited states. 
Thus,  the Boltzmann factor $\exp(-\beta E_n)$ acts as an (energy) eigenvalue filter, 
where large values of $\beta$ preferentially sample the low-energy part of the spectrum, and $\beta=0$ gives equal weight to the whole spectrum.

The SFF admits several information theoretic interpretations. In particular, it can be expressed as the fidelity \cite{Papadodimas15,delcampo17,delCampo2018,Xu21SFF} between a coherent Gibbs state 
$|\psi_\beta\rangle=\frac{1}{\sqrt{Z(\beta)}}\sum_ne^{-\beta E_n/2}|n\rangle$ and its unitary time-evolution 
\beqa
{\rm SFF}(t)=\left|\langle\psi_\beta|e^{-it H}|\psi_\beta\rangle\right|^2,
\label{cSFF}
\eeqa
or equivalently, as the survival probability of the evolving coherent Gibbs state. 
Likewise, in bipartite systems, it is convenient to consider the entangled state
\beqa
|{\rm TFD}\rangle=\frac{1}{\sqrt{Z(\beta)}}\sum_ne^{-\beta E_n/2}|n\rangle\otimes |n\rangle ,
\eeqa
known as the thermofield double state (TFD). 
In terms of it, ${\rm SFF}(t)= | \langle {\rm TFD}|e^{-it H\otimes\mathbb{I}}|{\rm TFD}\rangle|^2$. The TFD is the purification of the thermal state of a single copy of the system, obtained by doubling the Hilbert space. 
The TFD was first introduced as a convenient reference state to extract thermal averages in field theory \cite{Umezawa82}. 
The TFD dynamics was used early on to model the ``hot'' thermal vacuum observed outside the horizon of a single radiating eternal black hole \cite{israel1976}.
In the context of the AdS/CFT correspondence, it describes an eternal two-sided black hole in AdS \cite{Maldacena03,Maldacena13}.
The SFF captures the survival probability of the TFD state under unitary time evolution \cite{Papadodimas15,delcampo17,delCampo2018,Xu21SFF}.
The conjecture that back holes are maximally chaotic \cite{MSS16} has led to a surge of activity in the study of the dynamical manifestations  of quantum chaos in the SFF \cite{Cotler17,Dyer17,delcampo17,Gharibyan18,Balasubramanian22,Erdmenger23}.

In theoretical and numerical studies, it is customary to average the SFF by considering a Hamiltonian ensemble, e.g., in random-matrix theory or in disordered systems.
In such a scenario, a property is said to be self-averaging when its estimate using  a typical member of the ensemble and the explicit average over the ensemble coincide. 
Self-averaging largely eases numerical studies in many-body systems, disposing of the need for Hamiltonian ensemble averages in characterizing the desirable property of the system.
However, the SFF is not self-averaging \cite{Prange97}.

 The structure of the SFF in the time domain is well-characterized \cite{Cotler17,Dyer17}: it exhibits a slope-dip-ramp-plateau structure, as shown in Fig.~\ref{SFFSAfig1}{\sf{\textbf{a}}}, that is manifested under averaging over disorder or a Hamiltonian ensemble.
 In the absence of averages,  erratic time-domain fluctuations appear, making it difficult to appreciate some of its features. An exception is the SFF computed using the gauge-gravity duality in the semiclassical approximation, where the erratic fluctuations are absent \cite{saad2019semiclassical}.
These fluctuations are sometimes referred to as  noise \cite{Prange97}, or quantum noise \cite{Barbon14}, terminology  to be distinguished from the standard one in the theory of open quantum systems \cite{Gardiner00}.
 Erratic wiggles in the time domain are a consequence of the discreteness of the energy spectrum and can be associated with  quantum coherence in the energy eigenbasis in the time evolution of the coherent Gibbs state or the TFD. Quantum noise is further responsible for the lack of self-averaging in the SFF. 
Fluctuations with respect to the signal do not cancel out upon averaging, e.g., over a Hamiltonian ensemble \cite{Leviandier86,Argaman93,Argaman93prl,Eckhardt95,Prange97}. This can be quantified by the finite value of the relative variance (RV)
\beqa
{\rm RV}(t) = \frac{\la {\rm SFF}^2(t) \ra - \la {\rm SFF}(t) \ra^2}{\la {\rm SFF}(t) \ra^2} \; ,
\label{eq:RV}
 \eeqa
 that does not vanish as the size of the Hilbert space is increased. 
The lack of self-averaging of the SFF and the survival probability was analytically shown for random matrices and disordered spin models~\cite{Schiulaz20,TorresHerrera20,torres-herrera2020a}.  It has been related to the zeroes of the partition function in the complex temperature plane, known as Fisher zeroes  \cite{bunin2023fisher}. This implies that no matter how large the system size is, an ensemble average is required, adding an extra layer of complexity to numerical studies, which are generally challenging due to the large Hilbert space involved in analyzing many-body quantum systems. 
 As an alternative to averaging over a Hamiltonian ensemble, 
numerical and analytical studies often resort to running averages over time that smear ${\rm SFF}(t)$ over intervals of time \cite{Leviandier86,Eckhardt95}. A yet different approach resorts to modifying the definition of the SFF restricting the Fourier transform of the two-point function over an energy window, or more generally, using a filter function over an energy or frequency band \cite{Hammerich89,Wall95,Mandelshtam97,Prange97,Gharibyan18}. 

In what follows, we build on the interpretation of the SFF as a fidelity between quantum states related by time evolution and show that suppressing the erratic wiggles implies the breakdown of unitarity in the dynamics. To this end, we reformulate as quantum operations described by a (nonunitary) quantum channel, 
the different approaches to reduce the time fluctuations in the SFF, such as ensemble averages, and to enforce self-averaging, such as filters in the energy and frequency domain.
For a particular class of filters, the resulting channels are of the mixed-unitary class, and the information lost due to the unitarity breaking can be recovered.

The paper is organized as follows. We review the structure of the unfiltered SFF in Sec. \ref{RevSFF}, and introduce the generalization of the SFF to arbitrary physical processes in Sec. \ref{SFFarb}, paving the way to the description of  filtering of the SFF in terms of nonunitary  quantum channels in Sec. \ref{SFFfilt}. Physical mechanisms associated with energy-dephasing and giving rise to different spectral filters are discussed in Sec. \ref{EDsec}. Section \ref{SFFLong} discusses the filtered SFF as a function of the system size, while Sec. \ref{InfoLR} focuses on information recovery under mixed-unitary quantum channels and the frequency filter deconvolution. Fundamental limits to quantum noise associated with the fidelity-based SFF are presented in Sec.  \ref{SFFmeas}. The relation between eigenvalue filters and Hermitian Hamiltonian deformations is discussed in Sec. \ref{EFHDef}. Time-continuous master equations for frequency filters are derived in Sec. \ref{MEFF} before closing with a discussion and conclusions.

\section{Features of the spectral form factor in an isolated chaotic quantum system}
\label{RevSFF}

We start by reviewing the well-known structure of the SFF for a chaotic system in isolation.
The (unfiltered) SFF averaged over a Hamiltonian ensemble can generally be written down in terms of different contributions. 
Invoking the annealed approximation, which replaces the average of a quotient by the ratio of the averages at high temperature, and in the absence of degeneracies in the energy spectrum,  one finds \cite{Cotler17,delcampo17}
\beqa
\la {\rm SFF}(t)\ra=\frac{1}{\la Z(\beta)\ra^2}\bigg[|\la Z(\beta+it)\ra|^2+g_c(\beta,t)+\la Z(2\beta)\ra\bigg].\nonumber\\
\eeqa
The first term in brackets is known as the disconnected part as it can be derived from the average density of states $\la \rho(E)\ra=\la \sum_n\delta(E-E_n)\ra$ (one-point function), as 
$$\la Z(\beta+it)\ra=\int dE\la \rho(E)\ra e^{-(\beta+it)E}.$$
The second term captures correlations among eigenvalues and is governed by the Fourier transform $g_c(\beta,t)$ of the connected two-level correlation function of the energy spectrum $\la \rho(E)\rho(E')\ra_c=\la \rho(E)\rho(E')\ra-\la \rho(E)\ra\la\rho(E')\ra$. Specifically, 
$$ g_c(\beta,t)=\int dEdE'\la \rho(E)\rho(E')\ra_c e^{-(\beta+it)E}e^{-(\beta-it)E'}.$$
The last term is constant and governs the long-time asymptotics.
The SFF reduces to unit value at $t=0$. 
The short time evolution gives rise to a parabolic decay $\la {\rm SFF}(t)\ra =1-\la\Delta H^2\ra t^2$ in the time scale fixed by the inverse of the energy fluctuations $\la\Delta H^2\ra=\int d E (E-\la E\ra)^2\la \rho(E)\ra$ and extends, forming a slope. 
This decay is governed by the disconnected part of the SFF.
In chaotic systems, the decay reaches a dip below the long-time asymptotics. The region where $\la {\rm SFF}(t) \ra < \la Z(2\beta)\ra / \la Z(\beta)\ra^2$ is known as correlation hole or dip~\cite{Leviandier86,Alhassid1992}.
The latter is followed by a ramp, governed by the eigenvalue correlations, and is thus a proxy for quantum chaos. 
The ramp extends from the dip time to the plateau time, at which it takes the constant value $\la {\rm SFF}\ra= \la Z(2\beta)\ra / \la Z(\beta)\ra^2$ in the annealed approximation, in the absence of degeneracies expected in chaotic systems.

\section{Spectral form factor in arbitrary physical processes}
\label{SFFarb}

The fidelity-based interpretation of the SFF can be leveraged to consider more general sorts of time evolution beyond the unitary case.
In particular, this makes it possible to generalize the SFF to non-Hermitian and open quantum systems characterized by nonunitary evolution \cite{Tameshtit1992,Xu21SFF,Cornelius21,MatsoukasRoubeas23,MatsoukasRoubeas23PQC,ZhouZhouZhang23}. This section introduces tools used to describe nonunitary evolution that will be employed in the explanations that  follow in the next sections.

Several generalizations of the SFF have been put forward when the dynamics is not unitary.
At variance with alternative proposals with a restricted domain of applicability \cite{LiProsenAmos21,VikramGalitski22}, the fidelity-based generalization of the SFF has the advantage of involving only the eigenvalue correlations that govern quantum dynamics and  applies to arbitrary physical processes. 
Provided that the evolution is described by a quantum channel $\Phi_t(\cdot)$ (i.e., a completely-positive and trace-preserving map), the fidelity-based SFF is given by \cite{Xu21SFF,Cornelius21,MatsoukasRoubeas23,MatsoukasRoubeas23PQC,ZhouZhouZhang23} 
\beqa
{\rm SFF}(t)=\la \psi_\beta|\Phi_t(|\psi_\beta\ra\la\psi_\beta|)|\psi_\beta\ra. 
\eeqa
An arbitrary quantum channel admits a Kraus representation, $\Phi_t(\rho_0)=\sum_{\alpha=1}^r K_\alpha \rho_0K_\alpha^\dag$, where $r$ is known as the Choi rank \cite{BP02}. The case of unitary evolution corresponds to the case of a single Kraus operator that equals the time evolution operator, i.e., $K(t)=K_1(t)=U(t)$ and $K_\alpha(t)=0$ for $\alpha>1$.
Given that Kraus operators need only obey the condition of adding up to the identity $\sum_\alpha K_\alpha^\dag K_\alpha=\mathbb{I}$ for the dynamics to be trace-preserving and that the Kraus decomposition involves  $1\leq r\leq d^2$ Kraus operators in a $d$-dimensional Hilbert space, it is apparent that the chaotic features of the SFF under unitary dynamics are generally suppressed under nonunitary time evolution.
As a result, quantum channels with a simple representation in the energy eigenbasis are singled out to study filtering in quantum chaos and self-averaging of the SFF. 

An important class of channels that will be of relevance in the following is that of mixed-unitary channels \cite{Watrous18}.
A channel $\Phi$ is a mixed-unitary channel if there is an alphabet $\Sigma$, a probability vector $p$ and a collection of unitaries $\{U_y:y\in\Sigma\}$ such that
\beqa
\Phi(\rho)=\sum_{y\in\Sigma}p(y)U_y\rho U_y^\dag.
\eeqa
The channel is thus a convex combination of unitaries.
This kind of quantum channel is unital and thus preserves the identity $\mathbb{I}$, i.e., $\Phi_t(\mathbb{I})=\mathbb{I}$.

The fidelity-based interpretation of the SFF extends to higher moments of the SFF. Indeed, given that the initial state $\rho_0=|\psi_\beta\ra\la\psi_\beta|$ is pure, the $k$-th moment reads 
\beqa
{\rm SFF}^k=\tr[\underbrace{\rho_0\rho_t\dots\rho_0\rho_t}_{\textrm{$k$  times}}]=\la\psi_\beta|\rho_t|\psi_\beta\ra^k.
\eeqa
The $k$-th moment can be associated with a Zeno sequence in which the time evolution is interrupted by sequential projective measurements onto the initial state. 
Therefore, the RV in Eq.~(\ref{eq:RV}) probes the degree of factorization of the time evolution in a sequence with $k=2$ in the presence of averaging.

     \begin{figure*}[ht]
\hspace*{-0.0 cm}
\includegraphics[scale=0.45]{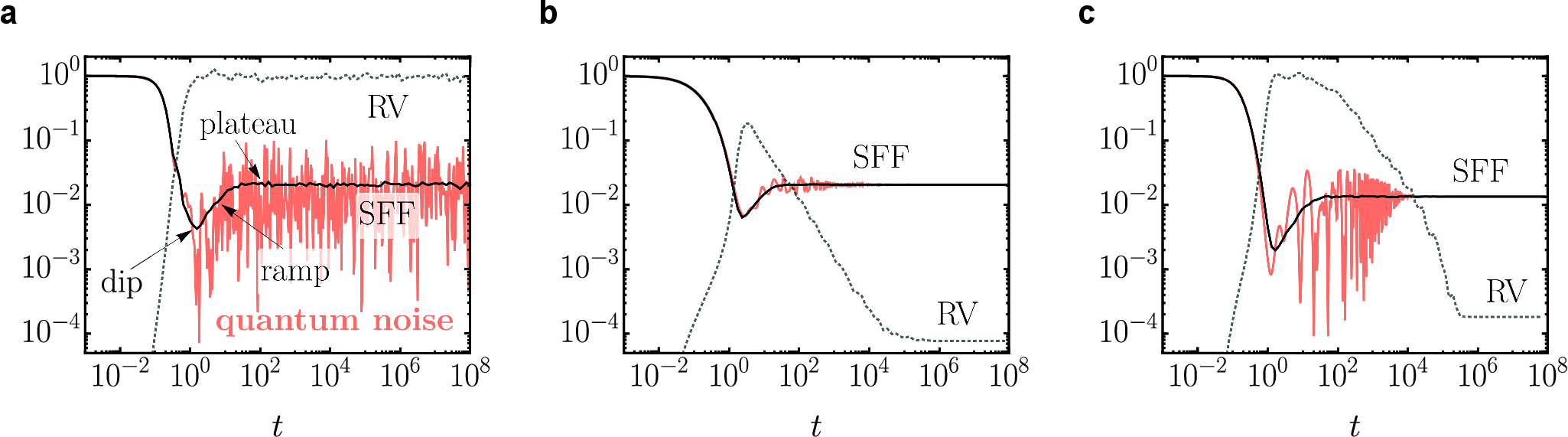}
  \caption{Spectral form factor for a single realization (solid red line) and upon Hamiltonian average (solid black line),  together with the corresponding RV (black dashed line). 
The averages are taken over a sample of $500$ random $\mathrm{GOE}(64)$ Hamiltonians $H$,  with $\sigma =1$.  
{\sf{\textbf{a}}} In the unfiltered case,  $\kappa=0$,  the RV saturates at the unit value after the dip time.  
{\sf{\textbf{b}}}  RV using frequency filtering with the Gaussian function \eqref{EDfilter} and a finite dephasing strength $\kappa=0.1$. 
The RV  reaches a maximum value at the dip time and then drops to a plateau of value  $\text{RV}_p = \big\langle ( Z(2 \beta)^2 / Z(\beta)^4 \big\rangle / \big\langle Z(2 \beta) / Z(\beta)^2 \big\rangle^2 $.
{\sf{\textbf{c}}} RV with eigenvalue filtering using the Gaussian function \eqref{BNGLfilter} with $f(E)=E$. 
The RV increases to its maximum at the dip time and then drops to a plateau given by  $\text{RV}_p = \big\langle  1 / Z(\beta)^2 \big\rangle / \big\langle 1 / Z(\beta) \big\rangle^2 $.  
In all three panels, the inverse temperature is  $\beta = 0.1$.      
}
  \label{SFFSAfig1}
    \end{figure*} 

\section{Unitarity breaking: spectrum filtering as a nonunitary quantum channel}
\label{SFFfilt}

In what follows, we consider three approaches frequently used to reduce the erratic wiggles in the SFF. They involve averaging over a Hamiltonian ensemble and the use of frequency filters and eigenvalue filters. The last two involve different kinds of time-averaging and ensure the self-averaging of the SFF at long times.  We show that all three cases involve unitary breaking described by a nonunitary quantum channel.

\subsection{Averaging over Hamiltonian ensembles}

Averaging over a Hamiltonian ensemble constitutes a popular approach that smooths out the quantum noise wiggles in the SFF.
This approach is at the core of the random-matrix theory, the study of disordered systems, and matrix models \cite{MethaBook,Haake,Forrester10}.
Given a Hilbert space $\mathcal{H}$ of dimension $d$, consider an ensemble of Hamiltonians $\mathcal{E}_{H}$ with a probability density function $P(H)$ and integration measure $dH$ such that $\int_{\mathcal{E}_{H}}P(H)dH=1$.
The average of the SFF over ${\mathcal{E}_{H}}$ is given by
\beqa
\la {\rm SFF}(t)\ra_{{\mathcal{E}_{H}}}=\int_{\mathcal{E}_{H}}dHP(H)\left|\frac{\tr\left(e^{-(\beta+it)H}\right)}{\tr\left(e^{-\beta H}\right)}\right|^2.
\eeqa
The fidelity-based interpretation of the SFF illuminates the underlying physical process involved in such an average.
For a specific Hamiltonian $H=\sum_nE_n|n\ra\la n|$, the initial state is chosen as the coherent Gibbs state (or the TFD in the case of a bipartite system),   
\beqa
|\psi_\beta(H)\ra=\sum_{n}\frac{e^{-\beta H/2}}{\sqrt{\tr\left(e^{-\beta H}\right)}}|n\ra.
\eeqa 
The Hamiltonian ensemble ${\mathcal{E}_{H}}$ provides an alphabet, together with the collection of unitaries $\{U_{H}(t)=e^{-itH}:H\in \mathcal{E}_{H}\}$. 
The state $|\psi_\beta(H)\ra$ is chosen with probability measure $P(H)dH$ and evolved unitarily into $U_{H}(t)|\psi_\beta(H)\ra$.
The SFF is then computed as the averaged survival probability over the Hamiltonian ensemble,
\beqa
\la {\rm SFF}(t)\ra_{{\mathcal{E}_{H}}}=\int_{\mathcal{E}_{H}}dHP(H)|\la \psi_\beta(H)|U_{H}(t)|\psi_\beta(H)\ra|^2.\nonumber\\
\eeqa
As a result, averaging the SFF over a Hamiltonian ensemble involves breaking the unitarity of the dynamics by classically mixing a distribution of states and unitaries. 
When the initial state $\rho_0$ is fixed and independent of the  Hamiltonian $H$, the  process can be associated with a mixed-unitary channel  
$
\Phi(\rho_0)=\int_{\mathcal{E}_{H}}dHP(H)U_{H}(t)\rho_0 U_{H}(t)^\dag.
$ 
As a relevant instance, this is the case when the initial state is the coherent Gibbs state in the infinite temperature limit $\beta=0$, $|\psi_\beta\ra=\sum_{n}\frac{1}{\sqrt{d}}|n\ra$, where the Hilbert space dimension fixes the probability amplitudes.

\subsection{Frequency filtering and the time-averaged SFF}
As an alternative to Hamiltonian averaging, 
in numerical and analytical studies, it is customary to 
 enforce the SFF's self-averaging by using a filter function $w(E_n-E_m)$ that acts on the frequency domain, suppressing contributions from given eigenvalue differences in the spectrum of a single Hamiltonian. This is equivalent to filtering eigenvalues of the Liouville superoperator $ \mathbb{L}=-i(H\otimes\mathbb{I}-\mathbb{I}\otimes H^T)$ that governs the unitary evolution in the vectorized density matrix $|\rho_t)$ according to $\frac{d}{dt}|\rho_t)= \mathbb{L}|\rho_t)$, i.e., when representing the Liouville-von Neumann equation as a linear matrix equation. We assume the frequency filter  to be described by a symmetric function $w(x):\mathbb{R}\rightarrow[0,1]$ satisfying $w(x)=w(-x)$. 
 The frequency-filtered SFF
is then proportional to
 $\sum_{nm}e^{-\beta (E_n+E_m)-it(E_n-E_m)}w(E_n-E_m)$.
 Making use of the Fourier transform of $w$, the frequency-filtered SFF reads
\beqa
\label{SFFFF}
{\rm SFF}_w(t) &=& \sum_{n,m=1}^{d}\frac{e^{-\beta (E_n+E_m)-it(E_n-E_m)}}{Z(\beta)^2}w(E_n-E_m)\nonumber\\
&=&\frac{1}{2\pi}\int_{-\infty}^{\infty} d\tau\widetilde{w}(t-\tau)\left|\frac{Z(\beta+i\tau)}{Z(\beta)}\right|^2,
\eeqa
with $\widetilde{w}(y)=\int_{-\infty}^{\infty} dE \exp(-iyE)w(E)$.
Filtering in frequency space is equivalent to time-averaging the canonical SFF associated with the unitary time evolution. Without degeneracies in the energy spectrum, the long-time behavior of ${\rm SFF}_w$ saturates at the plateau value set by $w(0)$.
 Further, in the fidelity-based interpretation of the SFF, frequency filtering can be recast as the result of a nonunitary time evolution. 
To this end, consider a quantum channel $\Phi_t$ such that the time-evolution of the initial coherent Gibbs state $|\psi_\beta\ra\la\psi_\beta|=\sum_{nm}e^{-\beta (E_n+E_m)/2}/Z(\beta)$ reads
\beqa
\label{rhoed}
\rho_t&=&\Phi_t(|\psi_\beta\ra\la\psi_\beta|)\\
&=&\sum_{nm}\frac{e^{-\beta (E_n+E_m)/2-it(E_n-E_m)}}{Z(\beta)}w(E_n-E_m)|n\ra\la m|.\nonumber
\eeqa
The latter can be rewritten as 
\beqa
\Phi_t(\rho_0)=\int_{-\infty}^{\infty} dy K(y)\rho_0K(y)^\dag,
\eeqa
with 
\beqa
K(y)=\left(\frac{\widetilde{w}(y)}{2\pi}\right)^{\frac{1}{2}}e^{-i(t+y)H}.
\eeqa
For the time evolution to be trace-preserving, it is required that
\beqa
\int_{-\infty}^{\infty} dy K(y)^\dag K(y)=\frac{1}{2\pi}\int_{-\infty}^{\infty} dy\widetilde{w}(y)=1,
\label{Knorm}
\eeqa
that is, $w(0)=1$.
The above equations provide an analog of the Kraus decomposition with a continuous index \cite{delCampo2020}. 
 They are associated with energy diffusion processes. 
Generally,  the Fourier transform  $\widetilde{w}(y)$ of the frequency filter can take both negative and positive values. 
However, given Eq.~\eqref{Knorm}, whenever $p(y)=\frac{1}{2\pi}\widetilde{w}(y)\geq 0$, it can be thought of as a probability distribution. 
Frequency filtering is then described by a mixed-unitary channel, i.e., the convex combination of unitary quantum channels, each with a single Kraus operator that equals the time-evolution operator shifted as $t\rightarrow t+y$. The collection of unitaries in this case  $\{U_y(t)=e^{-i(t+y)H}:y\in\mathbb{R}\}$ is generated by one single Hamiltonian $H$, leading to a time-average of the quantum state at time $t$, $\overline{\rho_t}=\int dy p(y)e^{-i(t+y)H}\rho_0 e^{i(t+y)H}$, from which the SFF is obtained as the fidelity ${\rm SFF}_w(t)=\la \psi_\beta|\overline{\rho_t}|\psi_\beta\ra$. 

An important example concerns the time averaging of the SFF over a time window of duration $T$, 
\beqa
\overline{\rm SFF}(t)=\frac{1}{T}\int_{-T/2}^{+T/2}\left|\frac{Z(\beta+it+iy)}{Z(\beta)}\right|^2 dy, 
\eeqa
for which   $\widetilde{w}(y)=2\pi/T$ for $y\in[-T/2,T/2]$ and zero otherwise. This is tantamount to considering the averaged time-dependent state
$\overline{\rho_t}=\frac{1}{T}\int_{-T/2}^{T/2} dy e^{-i(t+y)H}\rho_0 e^{i(t+y)H}$.

\subsection{Eigenvalue filtering}
An alternative filtering of the SFF involves expressions of the form $|\sum_{n}e^{-\beta E_n-itE_n}w(E_n)|^2$ with a filter function $w(E)\geq 0$ that acts directly on the eigenvalues. 
This is equivalent to selecting an energy band to study the SFF, while disregarding contributions from other parts of the spectrum \cite{Eckhardt95,Prange97}.
As noted in the introduction, the Boltzmann factor $\exp(-\beta E_n)$ can be considered as an exponential eigenvalue filter acting on the SFF with $\beta=0$. 
The use of an energy-eigenvalue filter function can be associated with the evolution governed by a single nonunitary Kraus operator
\beqa
K(t)&=&
e^{-itH}\sqrt{w}(H).
\eeqa
 
 The selection of the energy window corresponds to a post-selection represented by the operation
 \beqa
| \psi_\beta\ra\la\psi_\beta|\rightarrow \rho_t=\frac{K(t)|\psi_\beta\ra\la\psi_\beta|K(t)^\dag}{ Z_w(\beta)},
 \eeqa
 which is always a pure and normalized state, including the state at $t=0$. Here, the modified partition function 
 \beqa
 \label{ZwEF}
 Z_w(\beta)&=&\tr[K(t)| \psi_\beta\ra\la\psi_\beta|K(t)^\dag]\nonumber\\
 &=&\tr[w(H)e^{-\beta H}].
 \eeqa
 This accounts for the correct normalization, so that the SFF  at all times $t\geq 0$  is still given as the Uhlmann fidelity ${\rm SFF}_w(t)=\tr(\rho_0\rho_t)$, i.e.,  the survival probability of the post-selected  coherent Gibbs state $\rho_0$ and its time evolution, 
\beqa
{\rm SFF}_w(t)=
\sum_{nm}e^{-\beta (E_n+E_m)-it(E_n-E_m)}\frac{w(E_n)w(E_m)}{Z_w(\beta)^2}. \nonumber\\
\label{SFFwEF}
\eeqa
The choice of the Kraus operator is nonlinear in the quantum state, as it is tailored for the initial coherent Gibbs state, i.e., $\tr[K(t)|\psi_\beta\ra\la\psi_\beta|K(t)^\dag]=1$, making (only) in this case the dynamics trace-preserving. While this scenario is not the standard one in the theory of open quantum systems, it admits a natural interpretation in terms of energy dephasing without quantum jumps, as discussed in \ref{SecNHBGL}.

For completeness, we note that in terms of the Fourier transform of $\widetilde{w}(y)=\int_{-\infty}^{\infty} dE \exp(-iyE)w(E)$ and the definition $p(y)=\widetilde{w}(y)/(2\pi)$, the filtered SFF can be found in terms of the analytically continued partition function
\beqa
{\rm SFF}_w(t)=\frac{1}{Z_w(\beta)^2}\left| \int_{-\infty}^{\infty} d y p(t-y)Z(\beta-iy)\right|^2.
\eeqa
Naturally, for $w(E)=1$, $Z_w(\beta)=Z(\beta)$, $p(t-y)=\delta(t-y)$, one recovers the canonical SFF in Eq.~\eqref{cSFF}.

Before moving forward, let us characterize the performance of frequency and energy filters in the SFF. We consider random matrix Hamiltonians as a paradigm of quantum chaos. We sample the Hamiltonian matrices $H$ from the Gaussian orthogonal ensemble,  $\mathrm{GOE}(d)$,  calculate the corresponding ${\rm SFF}(t)$ and ${\rm SFF}^2(t)$, and then perform the average over the different realizations. 
Specifically,  we consider samples of real matrices $H=(X+X^\intercal)/2$,  where all elements $x \in \mathbb{R}$ of $X$ are pseudo-randomly generated with probability measure given by the Gaussian,  $\exp( -x^2/ \big(2 \sigma^2\big)) / ( \sigma \sqrt{2 \pi} )$,  where $\sigma$ is the standard deviation. 

Figure \ref{SFFSAfig1} shows three panels corresponding to the isolated, unfiltered SFF in panel {\sf{\textbf{a}}} and its modified versions with frequency and energy filters in panels {\sf{\textbf{b}}} and {\sf{\textbf{c}}}, respectively.
A single realization of the SFF exhibits quantum noise, manifested in the erratic oscillatory behavior in the time evolution (red line in Fig.~\ref{SFFSAfig1}{\sf{\textbf{a}}}). This is suppressed by performing a Hamiltonian ensemble average (solid black line in Fig.~\ref{SFFSAfig1}{\sf{\textbf{a}}}).
Alternatively, the frequency filter can suppress quantum noise in the SFF for a single random-matrix Hamiltonian without relying on ensemble averages, as illustrated in panel {\bf b}. Its effect is to reduce the oscillatory wiggles and the  RV. The use of filters acting on energy eigenvalues directly provides a different alternative, shown in panel {\bf c}. Note that for the unfiltered SFF, RV equals 1 from the time of the dip  onward, as seen in Fig.~\ref{SFFSAfig1}{\sf{\textbf{a}}}. This result holds for random matrices of any dimension~\cite{Schiulaz20} and for chaotic many-body quantum systems of any size~\cite{Schiulaz20,TorresHerrera20}, which means that the unfiltered SFF is non-self-averaging.  $\text{RV}=1$, because the distribution of the SFF$(t)$ for large times~\cite{torres-herrera2020a} is exponential, so the square of the mean of the distribution and its variance are equal. In contrast, the asymptotic values of the RV under frequency and energy filters become smaller than 1. Furthermore, as we shall see in Sec.~\ref{SFFLong}, the long-time values of the RV of the filtered SFF further decrease as $d$ increases, indicating that the SFF becomes self-averaging.

\section{Energy dephasing processes and spectral filtering}
\label{EDsec}

This section explores the relationship between energy-dephasing processes and the effects of different spectral filters.

\subsection{Frequency filters from energy dephasing}

Energy dephasing processes, also known as energy diffusion processes, arise in various scenarios \cite{Gisin84,Adler03}.
They are postulated in modifications of quantum mechanics involving wavefunction collapse models \cite{Milburn91,Bassi03,Bassi13}. They also arise in the description of unitary time evolution timed by a realistic clock subject to errors \cite{Egusquiza99,Egusquiza03}.
They have been used to study the interplay between quantum chaos and decoherence \cite{Xu19,Xu21SFF,Cornelius21}. Energy dephasing has also been analyzed in the context of AdS/CFT \cite{delCampo2020,Verlinde20,Verlinde21,Goto21} to explore the relation between entanglement and spacetime connectedness  \cite{Maldacena03}.
It can be described by the master equation 
\beqa
\label{MEED}
d_t\rho_t=-i[H,\rho_t]-\kappa[X,[X,\rho_t]],
\eeqa
with the condition that $[H,X]=0$, so that both Hermitian operators have a common set of eigenvectors, i.e., $H=\sum_nE_n|n\ra\la n|$ and $X=\sum_nx_n|n\ra \la n|$. The nested commutator plays the role of the dissipator and induces dephasing, suppressing coherent quantum superpositions in the energy eigenbasis. This is explicitly seen by considering the time evolution of an initial quantum state $\rho_0=\sum_{nm}\rho_{nm}(0)|n\ra\la m|$,
\beqa
\rho_t=\sum_{nm}\rho_{nm}(0)e^{-it(E_n-E_m)-\kappa t(x_n-x_m)^2}|n\ra\la m|.
\eeqa
For an initial coherent Gibbs state, the SFF is obtained as the survival probability
\beqa
\label{SFFED}
{\rm SFF}(t)=\sum_{nm}\frac{e^{-\beta (E_n+E_m)-it(E_n-E_m)}}{Z(\beta)^2}e^{-\kappa t(x_n-x_m)^2}.
\eeqa
When the Hermitian Lindblad operator is a deformation of the Hamiltonian, $X=f(H)$, $x_n=f(E_n)$, and $w(E_n-E_m)=\exp\{-\kappa t[f(E_n)-f(E_m)]^2\}$ in Eq. (\ref{SFFFF}).
When they are equal, $X=H$, one recovers the canonical case of energy dephasing.
In this case, one can recast ${\rm SFF}(t)$ in Eq.~\eqref{SFFED} as the frequency-filtered ${\rm SFF}_w$ \eqref{SFFFF} with the identification of a time-dependent Gaussian filter function
\beqa \label{EDfilter}
w(E_n-E_m)=\exp[-\kappa t(E_n-E_m)^2].
\eeqa

     \begin{figure}[t]
\hspace*{-0.2 cm}
\includegraphics[scale=0.45]{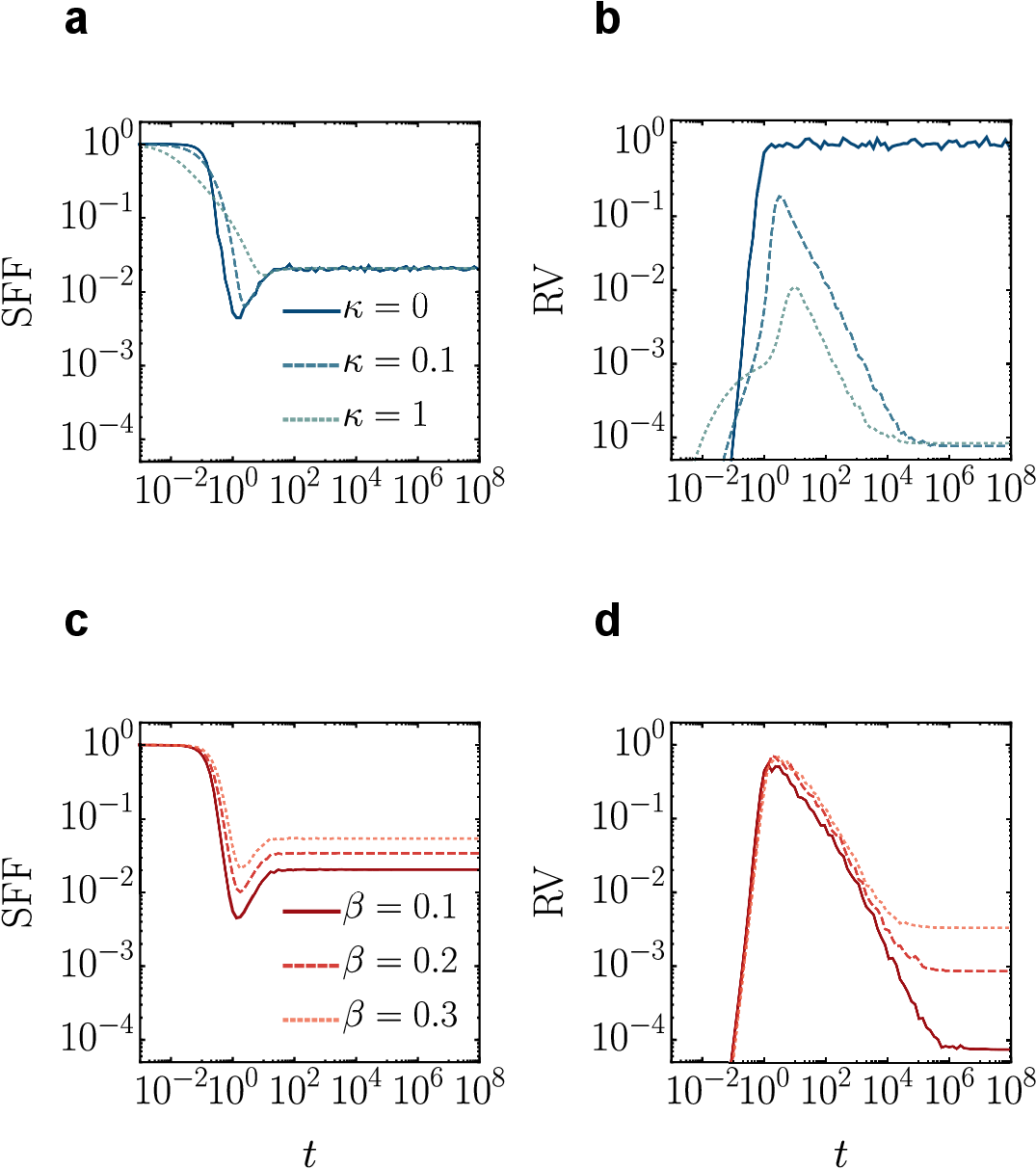}
  \caption{Frequency filtered SFF and its RV for different  dephasing strengths and inverse temperatures. 
Panels {\sf{\textbf{a}}} and {\sf{\textbf{b}}} show the SFF next to the corresponding RV for inverse temperature $\beta=0.1$ and different dephasing strengths $\kappa$.
Panels {\sf{\textbf{c}}} and {\sf{\textbf{d}}} show the SFF next to the RV for a dephasing strength $\kappa=0.01$ and different values of the inverse temperature $\beta$.
In all panels, the Hamiltonian averages were taken over a sample of $500$ random $\mathrm{GOE}(64)$ Hamiltonians $H$  with $\sigma =1$.
 }
  \label{SFFSAfig2}
    \end{figure} 

The action of the frequency filter \eqref{EDfilter} in the SFF is shown in Fig.~\ref{SFFSAfig2}{\sf{\textbf{a}}} for fixed $\beta=0.1$ and varying $\kappa$; see also Fig.~\ref{SFFSAfig1}. Such filtering delays the onset of the ramp, reduces its span, and decreases the depth of the correlation hole. In short, it decreases the dynamical manifestations of quantum chaos. 
The corresponding RV is shown in Fig.~\ref{SFFSAfig2}{\sf{\textbf{b}}} indicating that the long-time plateau of the RV is independent of $\kappa$ for $\kappa>0$, as expected from Eq.~\eqref{SFFED}. 
Figure~\ref{SFFSAfig2}{\sf{\textbf{c}}} shows the effect of varying $\beta$ for fixed $\kappa$, with the corresponding $\beta$-dependent long-time plateau being associated with the RV of a canonical thermal equilibrium state, as shown in Fig.~\ref{SFFSAfig2}{\sf{\textbf{d}}}.

\subsection{Eigenvalue filtering from energy dephasing without quantum jumps}
\label{SecNHBGL}

In what follows, we show that eigenvalue filtering can be described as the non-Hermitian evolution associated with energy-dephasing processes without quantum jumps.
To this end, consider the evolution operator $U(t)=\exp(-itH_T)$ generated by the time-independent non-Hermitian Hamiltonian $H_T=H-i\Gamma$, with $H=H^\dag$ and $\Gamma=\Gamma^\dag$.
In this case, the evolution is not trace-preserving, and one can introduce a single nonlinear Kraus operator dependent on the initial state $\rho_0$
\beqa
K=\frac{1}{\sqrt{\tr(e^{-itH_T}\rho_0e^{itH_T^\dag})}} e^{-itH_T}.
\eeqa
The latter is associated with a master equation of the form 
\beqa
d_t\rho_t=-i(H_T\rho_t-\rho_tH_T^\dag)+2\tr(\Gamma \rho_t)\rho_t,
\label{NLSE}
\eeqa
 which describes non-Hermitian dynamics subject to  balanced norm gain and loss \cite{Carmichael09,BrodyGraefe12}.

In particular, consider a non-Hermitian Hamiltonian in which the Hermitian and anti-Hermitian parts commute $[H,\Gamma]=0$ and thus have common eigenstates 
$\{|E_n\ra\}$. The action of the filter function can be identified by noting that $w(H)=\exp[-it\Gamma]$, i.e., $\Gamma|E_n\ra=-\frac{1}{t}\log w(E_n)|E_n\ra$.

As an illustrative example, 
consider the master equation for energy dephasing in Eq.~\eqref{MEED} with the condition $[H,X]=0$.  
This evolution is of the Lindblad form with a single Hermitian Lindblad operator $\sqrt{2}X$ and is thus Markovian \cite{BP02}.
As such, it can alternatively be written in terms of a non-Hermitian Hamiltonian $H_T=H-i2\kappa
X^2$ and a quantum jump term $\mathcal{J}(\rho)=2\kappa X\rho X$.
Disregarding the quantum jump term induces a non-Hermitian evolution exclusively governed by $H_T$. This can be justified at short times or by post-selection of quantum trajectories to the absence of quantum jumps \cite{Ashida20}. The evolution of the subset of quantum trajectories exhibiting no quantum jumps from time $t=0$ to time $t$ is governed by Eq.~\eqref{NLSE}, which is known as the nonlinear Schr\"odinger equation for null-measurement conditioning in this context \cite{Carmichael09}. Specifically, the time evolution subject to energy dephasing in the absence of quantum jumps is governed by \eqref{NLSE}, which admits a closed-form solution \cite{Cornelius21}. Explicit computation of the survival probability for the coherent Gibbs state yields the expression of the SFF
\beqa
{\rm SFF}(t)=\sum_{nm}e^{-\beta (E_n+E_m)-it(E_n-E_m)}\frac{e^{-\kappa t (x_n^2+x_m^2)}}{Z(\beta)Z_w(\beta,t)},\nonumber\\
\eeqa
where the modified partition function $Z_w(\beta,t)=\tr[w(X)^2\exp(-\beta H)]$.
The case of the Hamiltonian deformation $X=f(H)$ corresponds to the choice of the time-dependent filter function
\beqa \label{BNGLfilter}
w(E_n)=\exp[-\kappa t f(E_n)^2] ,
\eeqa
 in Eq.~\eqref{SFFwEF}.

 \begin{figure}[t]
\hspace*{-0.2 cm}
\includegraphics[scale=0.45]{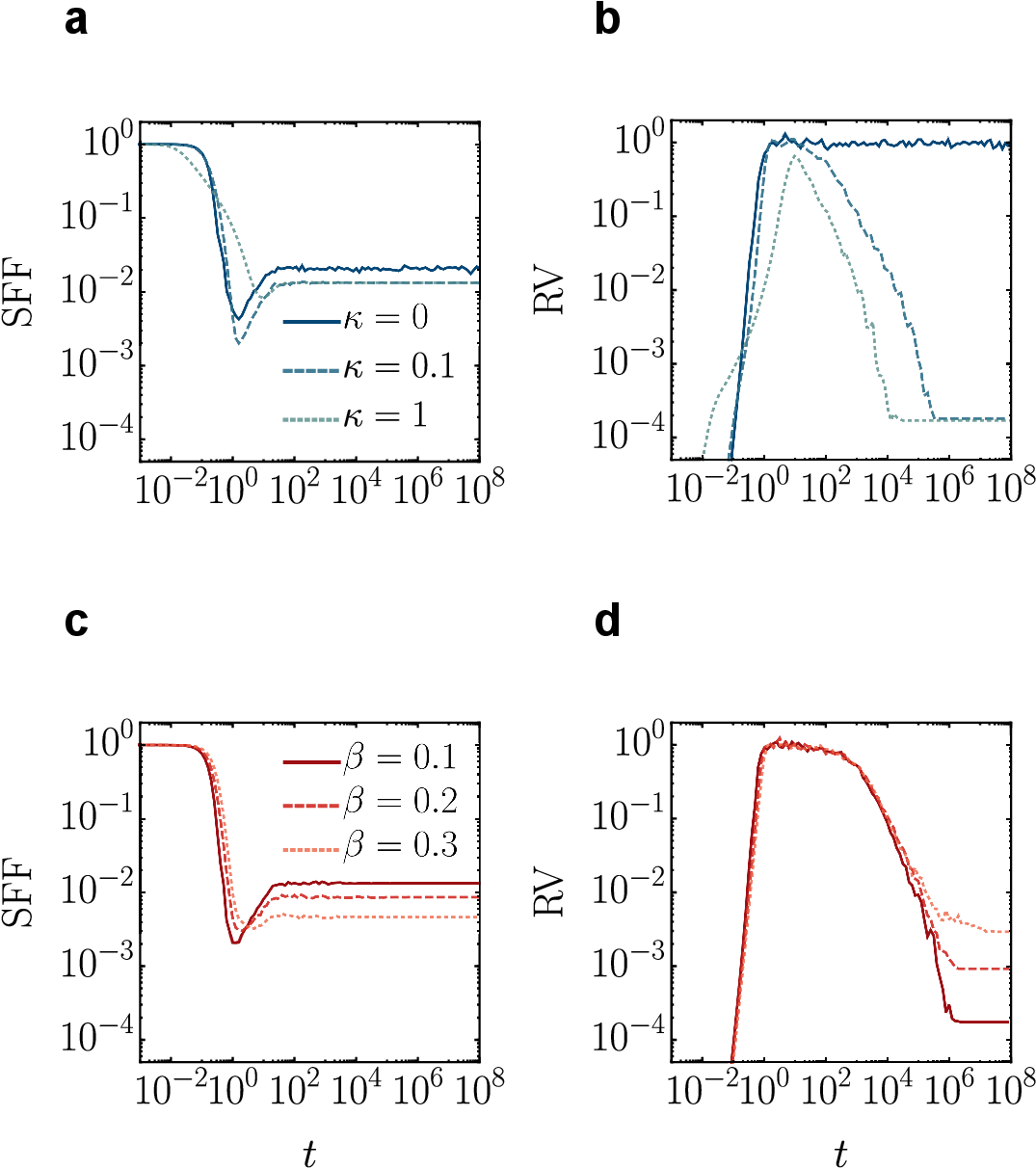}
 \caption{Eigenvalue filtered SFF and its RV for different dephasing strengths and inverse temperatures. 
Hamiltonian averages over a sample of $500$ random $\mathrm{GOE}(64)$ Hamiltonians $H$  with $\sigma =1$.
Panels {\sf{\textbf{a}}} and {\sf{\textbf{b}}} show the SFF and the corresponding RV with inverse temperature $\beta=0.1$ and different dephasing strengths $\kappa$.
Panels {\sf{\textbf{c}}} and {\sf{\textbf{d}}}  show the SFF and the associated RV with fixed dephasing strength $\kappa=0.01$ and varying inverse temperatures $\beta$.
 }
  \label{SFFSAfig3}
    \end{figure} 
The time dependence of the SFF with eigenvalue filtering in Eq. (\ref{BNGLfilter}) engineered through energy dephasing in the absence of quantum jumps is illustrated in Fig. \ref{SFFSAfig3}. At fixed $\beta$, increasing $\kappa$ reduces the correlation hole; see Fig.~\ref{SFFSAfig3}{\sf{\textbf{a}}}. For $\kappa>0$, the long-time plateaus of the SFF and RV  differ from the unfiltered case. Increasing $\beta$ for fixed $\kappa$ favors contributions to the SFF from the low-energy part of the spectrum and generally reduces the correlation hole and increases the plateau value of the SFF and the RV, as shown in Figs. \ref{SFFSAfig3}{\sf{\textbf{c}}}  and {\sf{\textbf{d}}}, respectively.

We emphasize that the definition of the SFF (\ref{SFFdef}) involves a finite inverse temperature $\beta$. In the fidelity-based interpretation, this presumes a TFD state with finite $\beta$ at $t=0$. The corresponding Boltzmann factors (probability amplitudes in the TFD) can be associated with an eigenvalue filter acting on an initial infinite-temperature TFD state.  
Varying the value of $\beta$ can be similarly associated with a non-Hermitian evolution conditioned to balanced norm gain and loss. 
We further notice that the difference in the SFF  at $\beta=0$ and  $\beta\rightarrow0^\pm$ has been associated with the emergence of many-body quantum chaos in a field theory analysis \cite{LiaoGalitski22}.

\section{Self-averaging at long times}
\label{SFFLong}

Under chaotic quantum dynamics,  quantities that are local in space are expected to be self-averaging at short times \cite{Schiulaz20,TorresHerrera20,torres-herrera2020a}.
It has  further been suggested that time-locality implies self-averaging at long times.
The SFF can be interpreted as a time auto-correlation function, thus a non-local quantity in time.
The unfiltered SFF lacks the self-averaging property at all timescales in isolated quantum systems \cite{Schiulaz20}.

We have shown that filters ubiquitously used to reduce the erratic wiggles of the SFF can be associated with quantum channels involving nonunitary dynamics. The breaking of unitarity contributes to suppressing quantum noise. 
In what follows, we numerically investigate the dependence of the RV as a function of the system size to identify when RV decreases as $d$ increases, thus rendering the SFF self-averaging.

\begin{figure}[t]
\hspace*{-0.2 cm}
\includegraphics[scale=0.43]{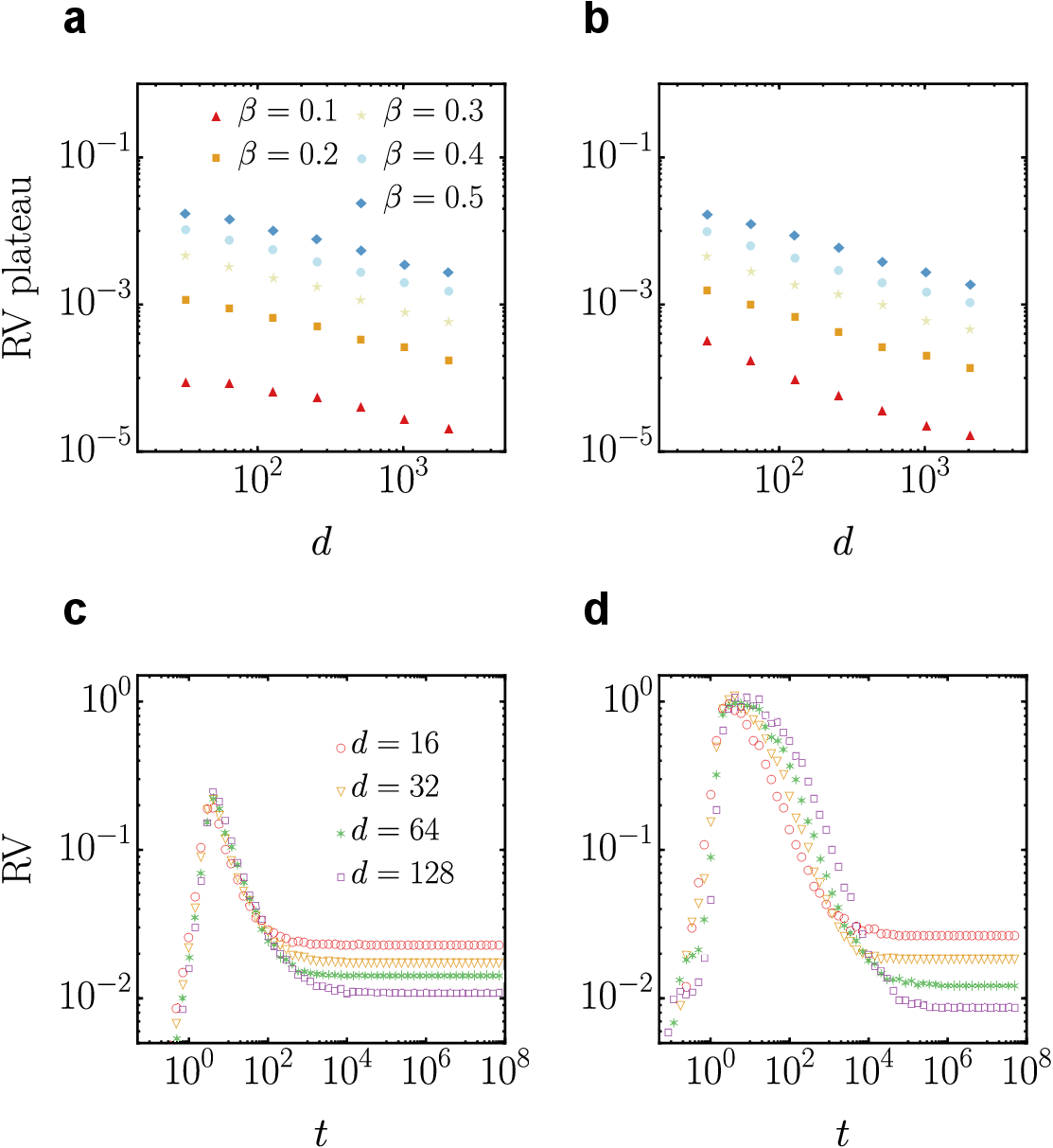}
  \caption{Asymptotic self-averaging of the filtered SFF.
Hamiltonian averages over a sample of $1000$ random $\mathrm{GOE}(d)$ Hamiltonians $H$  with $\sigma =1$.
Panels {\sf{\textbf{a}}} and {\sf{\textbf{b}}} show the plateau value of the frequency-filtered and the energy-filtered RV, that is independent of $\kappa$, as a function of the Hilbert space dimension $d$ is shown for different inverse temperatures.
Panel {\sf{\textbf{c}}} shows the frequency-filtered RV for inverse temperature $\beta=0.5$,  dephasing strength $\kappa=0.2$, and different Hilbert space dimensions $d$.
In panel {\sf{\textbf{d}}},  the corresponding energy-filtered RV for the same parameters is shown.
In both cases, the relative variance plateau decreases with the dimension increment, i.e.,  the SFF becomes self-averaging.
 }
  \label{SFFSAfig4}
    \end{figure}

Figure~\ref{SFFSAfig1}  implies that unitarity breaking can suppress the quantum noise of the SFF. 
Nevertheless,   the robustness against sample-to-sample fluctuations is associated with the reduction of the RV as the Hilbert space dimension increases.
Fig.~\ref{SFFSAfig4}{\sf{\textbf{a}}} and {\sf{\textbf{b}}} show that the frequency- and eigenvalue-filtered SFFs become self-averaging at the small inverse temperature shown and large times. Figure~\ref{SFFSAfig4}{\sf{\textbf{c}}}-{\sf{\textbf{d}}} confirm that the filtered SFFs become self-averaging at times after the correlation hole, where, according to Fig.~\ref{SFFSAfig1}{\sf{\textbf{b}}}-{\sf{\textbf{c}}}, $\langle \text{SFF}(t) \rangle >\text{RV}(t)$.   

\begin{figure}[t]
\hspace*{-0.2 cm}
\includegraphics[scale=0.43]{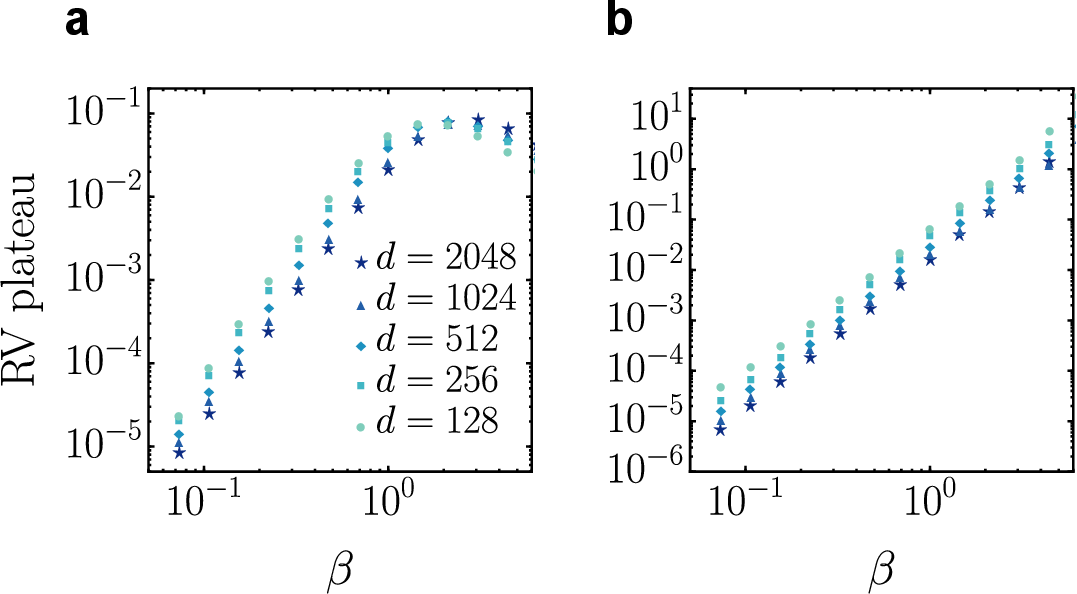}
  \caption{Self-averaging of the filtered SFF as a function of the inverse temperature. Panel {\sf{\textbf{a}}}  shows the value of the long-time RV plateau in the frequency-filtered SFF, reflecting a breakdown of self-averaging as the  inverse temperature is increased.  By contrast,  panel {\sf{\textbf{b}}} indicates that self-averaging remains robust against variations of the inverse temperature in the case  of eigenvalue filtering.}
  \label{SFFSAfig5}
    \end{figure}

The effect of the inverse temperature depends on the filter considered, as shown in Fig.~\ref{SFFSAfig5}. The long-time SFF is only self-averaging for moderate to high temperatures in the case of frequency filtering; see panel {\sf{\textbf{a}}}. By contrast, the long-time eigenvalue-filtered SFF remains self-averaging as the inverse temperature varies, as shown in panel {\sf{\textbf{b}}}.

\section{Information loss and its recovery}\label{InfoLR}

We have shown that the different approaches to suppress quantum noise in the SFF can be described as quantum channels involving nonunitary physical processes.
In particular, Hamiltonian averaging, frequency filtering, and time averaging of the SFF are all associated with mixed-unitary channels. The latter are unital and thus satisfy the necessary conditions for the purity $P_t=\tr[\Phi_t(\rho_0)^2]$ of the time-evolving state to decay monotonically under the action of the channel  \cite{Lidar06,Xu19}. Conversely, the linear entropy $S_L=1- P_t$ increases monotonically. Thus, these channels lead to monotonic information loss. Yet, the lost information is fully recoverable \cite{Gregoratti03,Watrous18}.
To appreciate this, it is convenient to consider the Hilbert space of the system together with the Hilbert space $\mathcal{H}_E$ of the environment with initial density matrix $\rho_E$, such that
\beqa
\Phi_t(\rho_0)=\tr_E\left(U_{SE}\rho_0\otimes\rho_E U_{SE}^\dag\right),
\eeqa
in terms of a global unitary $U_{SE}$. 
One can consider a measurement on the environment associated with a family of operators $M_y$, such that $\sum_yM_y=\mathbb{I}_E$.
The expectation value of an operator $A$ on the system can be described in terms of a family of completely positive maps $\Phi_y$.  
\beqa
\tr[\Phi_t(\rho_0)A\otimes \mathbb{I}_E]&=&\sum_y\tr_E\left(U_{SE}\rho_0\otimes\rho_E U_{SE}^\dag A\otimes M_y\right)\nonumber\\
&=&\sum_y\tr[\Phi_y(\rho_0)A].
\eeqa
The decomposition of the channel $\Phi_t=\sum_y\Phi_y$ is known as an instrument. The measurement of $M$ on the environment yields outcome $y$ and the quantum state $\Phi_y(\rho_0)/\tr[\Phi_y(\rho_0)]$ with probability $p(y)=\tr[\Phi_y(\rho_0)]$. 
It is then possible to select the reverse operation
\beqa
R_y=U_y^\dag \Phi_t(\rho_0) U_y,
\eeqa
so that the information-recovery channel is 
\beqa
R=\sum_yR_y\circ \Phi_y.
\eeqa
In short, the information acquired by performing a measurement on the environment can be used to reverse the action of the quantum channel $\Phi_t$ on the system, thus recovering the initial state.

This information-recovery protocol involves access to the degrees of freedom of an environment, which may be physical or an auxiliary construction, depending on the context. Any physical system is embedded in an environment that may give rise to decoherence and filtering through interaction with the system of interest. 
By contrast, in an effectively isolated system, one may still consider using nonunitary operations for filtering as done in numerical analysis without an explicit physical environment.  


In what follows, we tackle a complementary problem, the recovery of information masked exclusively by the filter.  We focus on frequency filtering and aim at obtaining the unfiltered SFF from the filtered one by undoing the action of the filter.
 The filtered SFF is the convolution of the Fourier transform of the filter function and the canonical SFF, as shown in Eq.~\eqref{SFFFF}, which can be written as
\beqa
{\rm SFF}_w(t)=\frac{1}{2\pi}\widetilde{w}(t)\ast{\rm SFF}(t).
\eeqa
By the convolution theorem, it is thus possible to retrieve ${\rm SFF}$ from knowledge of ${\rm SFF}_w$ and $w$, using
\beqa
\widetilde{{\rm SFF}}(\nu)=\frac{\widetilde{{\rm SFF}}_w(\nu)}{w(\nu)},
\eeqa
provided that $w(\nu)$ is nonzero everywhere in the domain of $\widetilde{{\rm SFF}}_w(\nu)$. 
Even when the inverse frequency filter function $1/w(\nu)$ is nonsingular, the inversion can be unstable for small values of $w(\nu)$. Furthermore, knowledge of $\widetilde{{\rm SFF}}_w(\nu)$  generally comes with additive noise, whether resulting from limited machine precision in a numerical simulation or statistical errors in measured data. This scenario is common in filter analysis and motivates alternatives to direct deterministic deconvolution, such as the Wiener deconvolution.

\section{Intrinsic quantum noise from eigenvalue statistics}\label{SFFmeas}

In the fidelity-based interpretation, the SFF is the survival probability of the time-evolving quantum state $\rho_t$ in the initial coherent Gibbs (or TFD) state.
As such, one can introduce a projector onto the initial state
\beqa
P=\rho_0=|\psi_\beta\ra\la\psi_\beta|, 
\eeqa
satisfying $P^2=P$, i.e., with eigenvalues $\pm1$. Such eigenvalues correspond to measurement outcomes in a projective measurement of $P$.  The full counting statistics associated with the projective measurement associated with $P$ is thus that of a discrete random variable, i.e., the Bernoulli distribution. Its characteristic function reads
\beqa
\tr\left[\rho_te^{i\theta P}\right]=1+(e^{i\theta}-1){\rm SFF}(t).
\eeqa
For any nontrivial evolution, an intrinsic quantum noise cannot be suppressed (other than by post-selection), whether the dynamics is unitary or not. The quantum noise associated with the uncertainty in the measurement outcomes of a projective measurement of $P$  can be quantified by the relative variance of the eigenvalue statistics encoded in the relation
\beqa
\frac{{\rm var}_{\rho_t}(P)}{\tr(P\rho_t)^2}=\frac{\tr(P^2\rho_t)-\tr(P\rho_t)^2}{\tr(P\rho_t)^2}=\frac{1-{\rm SFF}}{{\rm SFF}}.
\eeqa
For any $t>0$, up to recurrences of zero measure \cite{Bocchieri57,Schulman78}, ${\rm var}_{\rho_t}(P)>0$.

\section{Eigenvalue filtering as Hamiltonian deformation}\label{EFHDef}

We first note the following identity for the modified partition function (\ref{ZwEF}) with an  eigenvalue filter $w(E)$, 
\beqa
Z_w(\beta)=\tr\left[e^{-\beta\left(H-\frac{1}{\beta}\log w(H)\right)}\right].
\eeqa
As a result,  $Z_w(\beta)$ can be understood as the standard partition function of the operator
\beqa
F_\beta=H-\frac{1}{\beta}\log w(H),
\eeqa
that describes a one-parameter family of Hermitian Hamiltonian deformations of $H$ \cite{Gross20,Gross20b}. Formally, this deformation takes the form of a Helmholtz free energy operator analogous to that introduced to bound the charging power of quantum batteries \cite{GarciaPintos20}.
In particular, the filter gives rise to the entropy (surprisal) term $S(H)=\log w(H)$. The eigenvalue-filtered SFF in Eq. (\ref{SFFwEF})
 is then
\beqa
{\rm SFF}_w(t)=\left|\frac{\tr\left(e^{-\beta F_\beta-itH}\right)}{\tr\left(e^{-\beta F_\beta}\right)}\right|^2.
\eeqa
At long times, in the absence of degeneracies,  ${\rm SFF}_w(t)$ tends to
\beqa
\overline{{\rm SFF}_w}=\frac{\tr\left(e^{-2\beta F_\beta}\right)}{\tr\left(e^{-\beta F_\beta}\right)^2}.
\eeqa
This expression is nothing but the purity $P[\rho_w(\beta)]=\tr[\rho_w(\beta)^2]$ of the canonical Gibbs thermal state 
$\rho_w(\beta)$ defined with respect to the deformed Hamiltonian, i.e., the free energy operator $F_\beta$, 
\beqa
\rho_w(\beta)=\frac{e^{-\beta F_\beta}}{Z_w(\beta)}.
\eeqa
Indeed, the asymptotic value of ${\rm SFF}_w(t)$ can be written in terms of the second R\'enyi entropy $S_2[\rho_w(\beta)]=-\log \tr[\rho_w(\beta)^2]$ as
\beqa
\overline{{\rm SFF}_w}=P[\rho_w(\beta)]=e^{-S_2[\rho_w(\beta)]}.
\eeqa
For an eigenvalue filter function $w(E):\mathbb{R}\rightarrow[0,1]$, $\overline{{\rm SFF}_w}\leq \overline{{\rm SFF}}$ and $S_2[\rho_w(\beta)]\geq S_2[\rho(\beta)]$, where $\rho(\beta)=\exp(-\beta H)/Z(\beta)$ is the canonical thermal state of the undeformed Hamiltonian.

\section{Master equations for frequency filters from Liouvillian deformation}\label{MEFF}

We next show that frequency filters are associated with a family of master equations that generalize the dynamics related to energy dephasing.
Consider the master equation in which time-evolution is generated by a Liouvillian $\mathbb{L}$,
\begin{align} \label{vecdyn}
\frac{d}{dt}|\rho_t) = \mathbb{L} |\rho_t) ,
\end{align}
where $|\rho_t)$ denotes the vectorized density matrix at time $t$. In terms of it, ${\rm SFF}(t)=(\rho_0|\rho_t)$. 
Formally,  equation (\ref{vecdyn}) is solved by $|\rho_t) =e^{\mathbb{L} t} |\rho_0) $.
We focus on the case in which the Liouvillian is diagonalizable, so that it admits a spectral decomposition of the form
$\mathbb{L} = \sum_\mu \lambda_\mu |\mu)(\tilde \mu|$ using a bi-orthogonal basis. 
Here,  $|\mu)$ and $(\tilde \mu|$ are the right and left eigenstates, respectively, with complex eigenvalue $\lambda_\mu$ \cite{BrodyJPA2013,Gyamfi20}.
We next consider the Liouvillian of the form
\beqa
\mathbb{L}(\cdot) =-i [H,\cdot],
\eeqa
associated with an isolated system with Hamiltonian $H$. Its spectrum is purely imaginary, and left and right eigenvectors coincide. 
Given a complex function $W(z):\mathbb{C}\rightarrow \mathbb{C}$ we define the associated  Liouvillian deformation
 $W(\mathbb{L})=\sum_n W(\lambda_\mu) |\mu)(\mu|$ \cite{MatsoukasRoubeas23}.
By specifying   the Liouvillian deformation in terms of the frequency filter function $w(x):\mathbb{R}\rightarrow[0,1]$ as 
\beqa
W(\mathbb{L})=\log w(i\mathbb{L}), 
\eeqa
we consider a physical process in which the initial, unfiltered coherent Gibbs state $|\psi_\beta\ra\la\psi_\beta|$ evolves into a generalization of the frequency-filtered time-dependent density matrix in Eq.~\eqref{rhoed}. Specifically, we consider the time evolution for $t\geq 0$  described by the time-dependent density matrix 
\beqa
\label{rhoedkick}
\rho_t
&=&\sum_{nm}\frac{e^{-\beta (E_n+E_m)/2-it(E_n-E_m)}}{Z(\beta)}e^{\chi(t)W(E_n-E_m)}|n\ra\la m|,\nonumber
\eeqa
where $\chi(t)$  is a real function satisfying $\chi(0)=0$ and  $W(E_n-E_m)=\la n|W(\mathbb{L})|m\ra $ are matrix elements in the Hamiltonian eigenbasis.
This evolution fulfills the master equation
\beqa
\frac{d}{dt} |\rho_t) = \left[\mathbb{L}+\dot{\chi}(t)W(\mathbb{L})\right] |\rho_t),
\label{meff1}
\eeqa
with the initial condition $\rho_0=|\psi_\beta\ra\la\psi_\beta|$ and $\dot{\chi}$ denotes the time-derivative of $\chi$.   While $\mathbb{L}$ is anti-Hermitian,  $W(\mathbb{L})$ is Hermitian. Thus, $W(\mathbb{L})$ breaks unitarity and can be identified as the dissipator in the master equation (\ref{meff1}). 
Given that the $W(z)=W(-z)$, its Taylor series expansion involves only even powers of $z$, i.e., $W(z)=\sum_{n=0}^\infty W^{(2n)}(0)z^{2n}/(2n!)$.
The master equation can be written as
\beqa
\frac{d}{dt}\rho_t=-i[H,\rho_t]+\dot{\chi}(t)\sum_{n=0}^\infty \frac{W^{(2n)}(0)}{(2n)!}{\rm ad}_{H}^{2n}\rho_t,
\label{meff2}
\eeqa
where the nested commutators in each term of the Taylor series have been written compactly in terms of the adjoint map ${\rm ad}_{X}Y=[X,Y]$, ${\rm ad}_{X}^2Y=[X,[X,Y]]$, etc.

The case of a time-independent frequency filter for $t>0$ is described by choosing $\chi(t)$ as the Heavisde step function,
\beqa
\chi(t)=\Theta(t),\qquad \dot{\chi}(t)=\delta(t).
\eeqa
The delta function $\delta(t)=\frac{d}{dt}\Theta(t)$ in the master equation is thus required for the frequency filter to be time-independent. Implementing this filter relies on a single kick with the dissipator $W(\mathbb{L})$.  

Naturally, for the conventional energy-dephasing frequency filter \eqref{EDfilter}, the master equations \eqref{meff1} and \eqref{meff2} truncate at ${\rm ad}_{H}^{2}\rho$ and reduce to \eqref{MEED} for the choice $\chi(t)=t$, $\dot{\chi}(t)=1$.

\section{Discussion and conclusions}
The lack of self-averaging in the SFF is tied to quantum noise, manifested by erratic wiggles in the time domain. 
Analytical and numerical studies of the SFF enforce the reduction of the wiggles by resorting to Hamiltonian ensembles, time-averaging, and spectral filters in the energy or frequency domain. 
Through scaling analysis of the relative variance of the SFF, we have shown that the frequency and energy filters ensure that the SFF becomes self-averaging at long times.

We have established that suppressing the erratic wiggles (quantum noise) in the SFF implies nonunitary dynamics characterized by information loss and decoherence.
Hamiltonian averaging, time-averaging, and frequency filters can be described by a mixed-unitary channel representing the application of a random unitary with a given probability distribution. Mixed-unitary channels are unital and induce information loss that can, however, be recovered by environment-assisted channel correction.
By contrast, filters acting directly in the energy eigenvalues can be interpreted as a nonlinear quantum channel describing the non-Hermitian evolution of an energy-dephasing process conditioned to the absence of quantum jumps.

The identification of the canonical, filtered SFFs for isolated systems in terms of the survival probability of a coherent Gibbs state under nonunitary evolution singles out the fidelity-based generalization of the SFF to open quantum systems put forward in Refs. \cite{Xu21SFF,Cornelius21,MatsoukasRoubeas23,MatsoukasRoubeas23PQC,ZhouZhouZhang23} with respect to alternative proposals \cite{LiProsenAmos21,VikramGalitski22}.
The fidelity-based approach makes it possible to unify SFFs  for isolated systems with and without filters and for open quantum systems in a single framework.
Further studies of self-averaging SFFs can be envisioned by tailoring the filter function in accordance with the system size or Hilbert space dimension.

Our results rely on the combination of tools in quantum information science and quantum chaos, and contribute to the understanding of filters in the characterization of the spectral properties of many-body systems (e.g., in numerical studies) as physical operations breaking unitary. In particular, our results establish how such filters can be implemented in digital or analog quantum simulation experiments of the nonequilibrium dynamics of many-body systems.
This conclusion should be generalizable to quantities other than the SFF, such as correlation functions, that admit an information-theoretic interpretation associated with a quantum evolution. 
Our results hold for the dynamics of finite-dimensional systems and thus can be applied to the description of black hole physics in this framework, where self-averaging SFFs appear in a semiclassical description. In view of our findings, the latter involves unitarity breaking. 

\acknowledgements
It is a pleasure to acknowledge discussions with
Federico Balducci, Aurelia Chenu, Julien Cornelius, \'I\~{n}igo L. Egusquiza, Pablo Mart\'inez-Azcona, Federico Roccati,  Avadh Saxena, and Zhenyu Xu. This project was funded within the QuantERA II Programme that has received funding from the European Union’s Horizon 2020 research and innovation programme under Grant Agreement No. 16434093. 
For open access and in fulfillment of the obligations arising from the grant agreement, the authors have applied a Creative Commons Attribution 4.0 International (CC BY 4.0) license to any Author Accepted Manuscript version arising from this submission.  L.F.S. was supported by a grant from the United States National Science Foundation (NSF, Grant No. DMR-1936006).

\bibliography{SFFSA}

\end{document}